\documentclass[11pt, a4paper]{article}

\usepackage{stolenstyle}

\usepackage{amsmath,epsf,amssymb,latexsym,amsthm,setspace,bbm,array}

\usepackage[matrix,arrow,frame,import,curve,color]{xy}

\DeclareFontFamily{U}{rsf}{}
\DeclareFontShape{U}{rsf}{m}{n}{
  <5> <6> rsfs5 <7> <8> <9> rsfs7 <10-> rsfs10}{}
\DeclareMathAlphabet\Scr{U}{rsf}{m}{n}

\def\CO#1#2{{[#1,#2]}}
\def\AC#1#2{{\{#1,#2\}}}

\def\cDb{{\overline{\cD}}}
\def\cQb{{\overline{\cQ}}}

\def\GUL{\GU(1)_{\text{L}}}
\def\GUR{\GU(1)_{\text{R}}}

\def\End{\operatorname{End}}

\def\ch{\operatorname{ch}}

\def\SO{\operatorname{SO}}

\def\SU{\operatorname{SU}}
\def\GU{\operatorname{U{}}}

\def\GE{\operatorname{E}}

\def\p{\partial}
\def\pb{\bar{\partial}}

\def\ff#1#2{{\textstyle\frac{#1}{#2}}}
\def\half{\frac{1}{2}}

\def\cA{{\cal A}}

\def\cC{{\cal C}}
\def\cD{{\cal D}}

\def\cF{{\cal F}}

\def\cH{{\cal H}}

\def\cK{{\cal K}}
\def\cL{{\cal L}}

\def\cO{{\cal O}}

\def\cQ{{\cal Q}}

\def\cV{{\cal V}}

\newcommand\alphab{\overline{\alpha}}
\newcommand\betab{\overline{\beta}}
\newcommand\gammab{\overline{\gamma}}

\newcommand\thetab{\overline{\theta}}

\newcommand\xib{\overline{\xi}}

\newcommand\phib{\overline{\phi}}

\newcommand\psib{\overline{\psi}}

\newcommand\lambdat{\widetilde{\lambda}}

\newcommand\Gammab{\overline{\Gamma}}

\newcommand\Phib{\overline{\Phi}}

\newcommand\Lambdat{\widetilde{\Lambda}}

\newcommand\cb{\overline{c}}

\newcommand\hb{\overline{h}}
\newcommand\ib{\overline{\imath}}
\newcommand\jb{\overline{\jmath}}
\newcommand\kb{\overline{k}}

\newcommand\mb{\overline{m}}

\newcommand\qb{\overline{q}}

\newcommand\zb{\overline{z}}

\newcommand\Gb{\overline{G}}

\def\cAb{\overline{\cA}}
\def\cDb{\overline{\cD}}

\def\cKb{\overline{\cK}}
\def\cQb{\overline{\cQ}}

\def\bQ{{{\boldsymbol{Q}}}}
\def\bQb{{{\overline{\bQ}}}}
\def\tac{{{\text{(a,c)}}}}
\def\tcc{{{\text{(c,c)}}}}

\def\preprint{ {{AEI-2011-070}}}
\title{On marginal deformations of (0,2) non-linear sigma models}
\author[a] {Ilarion V.~Melnikov}
\author[b] {and Eric Sharpe}
\affiliation[a]{Max-Planck-Institut f\"ur Gravitationsphysik (Albert-Einstein-Institut)\\
 Am M\"uhlenberg 1, D-14476 Golm, Germany}
\affiliation[b]{Physics Department\\ Virginia Tech \\ Blacksburg, VA 24061, USA}
\emailAdd{ilarion.melnikov@aei.mpg.de}
\emailAdd{ersharpe@vt.edu}
\abstract{An N=1, d=4 supersymmetric compactification of the perturbative heterotic
string is described by a d=2 (0,2) superconformal field theory.  The first-order marginal deformations of the 
internal (0,2) SCFT are in 1 to 1 correspondence with massless gauge-neutral scalars in the
spacetime theory.  Working at tree-level in the $\alpha'$ expansion, we describe these first order
deformations for SCFTs with a (0,2) non-linear sigma model description.  Our results clarify the structure of deformations of heterotic Calabi-Yau compactifications and more general  
heterotic flux vacua.}

\begin{document}

\maketitle

\section{Introduction}
A perturbative heterotic string compactification that preserves $N=1$ super-Poincar\'e
invariance in four dimensions has a worldsheet description as a 
unitary (0,2) superconformal field theory (SCFT) with integral R-charges~\cite{Banks:1987cy} orbifolded by a heterotic GSO projection.  The resulting massless spectrum consists of the the minimal supergravity multiplet, the axio-dilaton chiral multiplet, 
vector multiplets for the spacetime gauge group $G$, ``matter'' chiral multiplets 
charged under $G$, as well as a number of $G$-neutral chiral multiplets.  The latter parametrize  $\cV$---the space of first order deformations of the (0,2) SCFT, which consists of right-chiral primary SCFT states with  conformal weights $(h,\hb) = (1,\half)$. 

A well-understood example is offered by a theory with (2,2)
worldsheet supersymmetry~\cite{Green:1987sp,Polchinski:1998rr}, in which case
 $\cV$ has a decomposition into three types of states
with respect to the (2,0) superconformal algebra:  $\cV_\tac$, $\cV_\tcc$,
and $\cV'$.  The first two are N=2 descendants of elements
of the (a,c) and (c,c) rings of the (2,2) SCFT, while $\cV'$ denotes any
additional (0,2) chiral primaries.  When the (2,2) theory is well-approximated
by a non-linear sigma model (NLSM) with a Calabi-Yau target-space $M$, the decomposition has a geometric interpretation in terms
of certain cohomology groups,   
leading to the familiar terminology of ``K\"ahler, complex structure, and bundle moduli.''
While useful on the (2,2) locus, the decomposition relies on
the accidental (2,0) supersymmetry, and in generic (0,2) theories the familiar
terminology becomes a less than useful misnomer.  This can be clearly seen in
F-theory constructions, where the ``bundle'' and ``complex structure'' deformations
enter on a more symmetric footing~\cite{Donagi:2009ra}.  In the heterotic context, first
order deformations  have been recently explored from the supergravity point of 
view in heterotic flux vacua~\cite{Becker:2006xp},  as well as in compactifications
involving a choice of a stable holomorphic bundle over a Calabi-Yau 
manifold~\cite{Anderson:2011ty}.

The aim of this note is to examine the space of first order deformations $\cV$ from the worldsheet point of view in the context of a (0,2) NLSM.  Working at tree-level in $\alpha'$ and using some simple (0,2) superspace techniques,  we will find a hands-on description of $\cV$.  While involving
ingredients familiar from the usual ``K\"ahler, complex structure, bundle'' decomposition,
and reducing to known results on the (2,2) locus, we will see that in general $\cV$ differs markedly from its (2,2) form. 

Our results agree with and generalize the supergravity analysis of heterotic Calabi-Yau compactifications.
Formally they also apply to heterotic flux vacua without a large radius Calabi-Yau limit.  To the extent that
the NLSM and geometry are good guides to such vacua,\footnote{One might expect this to hold in
vacua with extended spacetime supersymmetry.} our results provide a starting point for describing the moduli
space of heterotic flux compactifications. 

The rest of the note is organized as follows:  in section~\ref{s:nlsm} we set up the
tree-level (0,2) NLSM; in section~\ref{s:def} we describe the first order deformations;
section~\ref{s:examples} is devoted to checking the analysis by comparing to known cases,
and we end with some concluding remarks in section~\ref{s:discuss}.

\acknowledgments 
IVM would like to thank Ido Adam and Ronen Plesser for a fruitful (but as yet unpublished) 
collaboration on aspects of (0,2) SCFT that have been very useful for the present work.
Some of this work was completed while IVM was visiting the Virginia Tech Center for Neutrino
Physics, and he would like to thank the group for their hospitality.  The work of ES is partially
supported by NSF grants PHY-0755614 and PHY-1068725.

\section{The (0,2) NLSM} \label{s:nlsm}
In this section we will review some basic properties of (0,2) NLSMs relevant for
heterotic compactification.  Throughout, the geometric setup will be a stable 
holomorphic bundle $E \to M$ satisfying the usual anomaly cancellation conditions
$\ch_2(E) = \ch_2(T_M)$, where $M$ is a Hermitian $3$-fold with trivial canonical bundle.
To be concrete, we will restrict attention to models with $G = G' \times\GE_8$ and
$c_1(E) = 0$.  These theories possess an additional structure on the worldsheet:
a non-anomalous left-moving $\GU(1)$ symmetry $\GUL$, and as in Gepner's original
construction~\cite{Gepner:1987vz}, the GSO projection ensures that the $\SO(k)$ gauge symmetry associated to $k$ free left-moving fermions combines with 
$\GUL$ to form $G'$.\footnote{The $k$ free fermions and the ``hidden'' $\GE_8$ current lead to a modular-invariant critical string.}  Since we will be interested in the gauge-neutral sector, we will from now on focus on the internal theory.  Apart from a few small details
of conventions, we are following the standard treatment, as reviewed 
in, e.g.~\cite{Distler:1995mi}.

\subsection{(0,2) Superspace and the NLSM Lagrangian}
 We work in Euclidean signature with (0,2) superspace coordinates $(z,\zb,\theta,\thetab)$, with covariant derivatives $\cD,\cDb$ and supercharges $\cQ,\cQb$ 
given by\footnote{Our conventions have the advantage of not being cluttered by
factors of $i$; however, the price to pay is a non-standard charge conjugation action on the fermions: $\cC(\gamma) = \gammab$, and $\cC(\gammab) = -\gamma$.}
\begin{align*}
\cD & = \frac{\p}{\p \theta} + \thetab \pb, & 
\cDb & = \frac{\p}{\p \thetab}  + \theta \pb, \nonumber\\
\cQ &  = - \frac{\p}{\p\theta} + \thetab \pb, &
\cQb & = - \frac{\p}{\p\thetab} + \theta \pb,
\end{align*}
where $\pb \equiv \p /\p \zb$.
The non-trivial anti-commutators are 
$$\AC{\cD}{\cDb} = +2 \pb \quad\text{and}\quad \AC{\cQ}{\cQb} = -2 \pb.$$
Note that $\cD$ and $\cQ$ have $\GUR$ charge $\qb=-1$, while $\cDb$ and $\cQb$ have $\qb=+1$. 
Our basic fields are the chiral matter and chiral Fermi fields,
\begin{align*}
\Phi = \phi+\sqrt{2} \theta\psi+\theta\thetab \pb \phi, \qquad
\Gamma = \gamma + \sqrt{2} \theta G + \theta\thetab \pb \gamma,
\end{align*}
as well as the anti-chiral conjugate fields
\begin{align*}
\Phib = \phib-\sqrt{2} \thetab\psib-\theta\thetab \pb \phib, \qquad
\Gammab = \gammab + \sqrt{2} \thetab \Gb - \theta\thetab \pb \gammab.
\end{align*}
By construction, $\cDb$ ($\cD$) annihilates the chiral (anti-chiral) fields.

To build the NLSM Lagrangian, we take $3$ chiral multiplets $\Phi^i$, $r$ Fermi multiplets $\Gamma^\beta$ and their conjugates.  Assuming the NLSM will describe a superconformal theory, each $\Phi$ ($\Gamma$) multiplet contributes $(2,3)$ ($(1,0)$) to the central charge $(c,\cb)$; furthermore the $\GUL$ symmetry can be taken to act just on $\Gamma$ and $\Gammab$, assigning charges $1$ and $-1$, respectively, while $\GUR$ symmetry leaves both $\Gamma$ and $\Phi$ invariant.  With these assumptions, the most general (0,2) supersymmetric
Lagrangian is 
\begin{align}
\label{eq:Lgen}
4\pi \alpha' \cL = \cD \cDb \left[ \ff{1}{2} (\cK_i(\Phi,\Phib) \p \Phi^i - \cKb_{\ib}(\Phi,\Phib) \p \Phib^{\ib}) - \cH_{\beta\alphab}(\Phi,\Phib) \Gammab^{\alphab} \Gamma^\beta\right].
\end{align}
Here $\cH_{\alpha\betab}(\Phi,\Phib)$ is a Hermitian metric on the fibers of the bundle $E\to X$, while $\cK_i$ and $\cKb_{\ib}$ satisfy a reality condition $(\cKb_{\ib} )^\ast = \cK_{\ib}.$
The $\cK_i$ should be thought of as a locally defined (1,0) form $\cK = \cK_i d\phi^i$, and the action is invariant under shifts $\delta\cK = \omega$ for any holomorphic (1,0) form $\omega$, as well as under $\delta\cK = i \p f$ for some real function $f(\phi,\phib)$.  In
addition, setting $\cH' = U \cH U^\dag$ for any unitary transformation $U$ leads to an equivalent theory.
The free action with canonically normalized fields corresponds to $\cK_i = \Phib^i$ and $\cH_{\beta\alphab} = \delta_{\beta\alphab}$.

\subsection{Equations of motion and component expansion}
The equations of motion following from~(\ref{eq:Lgen}) can be derived by two well-known results:  first, if $X$ is a general (0,2) superfield, then
$$\cD\cDb(A X) |_{\theta,\thetab =0} = 0 \quad \forall X \implies A = 0;$$
second, any chiral (anti-chiral) superfield, say $\delta\Phi^i$ ($\delta \Phib^i$), can be expressed as $\cDb X$ ($\cD X$) for some general superfield $X$.  Varying the action in (\ref{eq:Lgen}), we obtain, up to total derivatives,
\begin{align*}
8\pi \alpha' \delta\cL & = \cD \cDb \Bigl[
\left\{ \cK_{i,j} \p\Phi^i - \p \cK_j -\cKb_{\ib,j} \p\Phib^{\ib}
-2 \cH_{\beta\alphab,j} \Gammab^{\alphab} \Gamma^\beta \right\} \delta\Phi^j
-2 \Gammab^{\alphab} \cH_{\beta\alphab} \delta\Gamma^\beta \Bigr]
+\text{h.c.} ~,
\end{align*}
which leads to the equations of motion
\begin{align}
\label{eq:eomraw}
0 = E^\Phi_j & = \cDb \left[ (\cK_{j,\ib}+\cKb_{\ib,j}) \p \Phib^{\ib}\right] 
+(\cK_{j,i\kb} -\cK_{i,j\kb}) \p\Phi^i \cDb \Phib^{\kb}
 +2\cDb(\cH_{\beta\alphab,j} \Gammab^{\alphab}) \Gamma^\beta, \nonumber\\
0 = E^\Gamma_\beta & = \cDb \left[ \cH_{\beta\alphab} \Gammab^{\alphab} \right].
\end{align}
The lowest component of $E^\Gamma_\beta$  and its conjugate yield the
equations of motion for the auxiliary fields $G$ and $\Gb$:
\begin{align*}
%\label{eq:Geom}
%\Gb^{\alphab}  = 
%- \cH^{\alphab \beta} \cH_{\beta\betab,\jb}  \gammab^{\betab}\psib^{\jb}, \qquad
%G^\alpha = \cH^{\betab\alpha} \cH_{\beta\betab,j} \gamma^\beta \psi^j.
\Gb^{\alphab}  = -\cAb^{\alphab}_{\betab\jb} \gammab^{\betab}\psib^{\jb}, \qquad
G^\alpha = \cA^\alpha_{\beta j} \gamma^\beta \psi^j,
\end{align*}
where $\cA$ and $\cAb$ denote components of the 
Hermitian connection on $E$ constructed from the metric $\cH$ and
its inverse:
\begin{align*}
%\label{eq:Adef}
\cA^{\alpha}_{\beta j} = \cH^{\betab\alpha} \cH_{\beta\betab,j}, \qquad
\cAb^{\alphab}_{\betab\jb} = \cH^{\alphab \beta} \cH_{\beta\betab,\jb}.
\end{align*}

With a little work we can also obtain the component expansion of the Lagrangian.
Up to boundary terms, we find
\begin{align}
\label{eq:fullS}
2\pi \alpha' \cL & =\ff{1}{2}g_{i\jb} (\p\phi^i\pb\phib^{\jb}+\p\phib^{\jb} \pb\phi^i)+
\ff{1}{2} B_{i\jb} (\p\phi^i\pb\phib^{\jb}-\p\phib^{\jb} \pb\phi^i) \nonumber\\
&\quad
+g_{i\jb} \psib^{\jb} \p\psi^i + \psib^{\ib} \left[
\p\phi^k \Omega^-_{\ib kj}
+\p\phib^{\kb}\Omega^-_{\ib\kb j} \right]\psi^j \nonumber\\
& \quad+ \gammab_{\mu} ( \pb\gamma^\mu + \pb\phi^j \cA^\mu_{\beta j} \gamma^\beta)
+\gammab_\mu \cF^\mu_{\beta\jb k} \gamma^\beta \psi^k\psib^{\jb},
\end{align}
where $\gammab_\mu \equiv \cH_{\mu\betab} \gammab^{\betab}$,  and
$\cF^\mu_{\beta\jb k}  = \cA^{\mu}_{\beta k,\jb}$ is the (1,1) component of the curvature for the connection $\cA$; the metric $g$
and B-field are given by
\begin{align*}
%\label{eq:gB}
g_{i\jb} = \ff{1}{2} (\cK_{i,\jb} +\cKb_{\jb,i}), \qquad
B_{i\jb} = \ff{1}{2} (\cK_{i,\jb}-\cKb_{\jb,i}),
\end{align*}
and $\Omega^-$ denotes the $H$-twisted connection 
\begin{align}
\label{eq:Omega}
\Omega^-_{\ib k j} = \Gamma_{\ib k j} - \ff{1}{2} H_{\ib k j} ,\qquad
\Omega^-_{\ib\kb j} = \Gamma_{\ib\kb j} -\ff{1}{2} H_{\ib\kb j},
\end{align}
where $H = d B$ is the tree-level torsion and $\Gamma$ is the Hermitian  Christoffel
connection for $g$.  As expected from the spacetime analysis~\cite{Strominger:1986uh}, the torsion is determined by the Hermitian form: $H_{ij\kb} = g_{\kb j,i}-g_{\kb i,j}$.  For what follows it will be useful to recast
the superspace equations of motion in terms of $g$ and $\Omega^-$:
\begin{align}
\label{eq:eom}
g_{i\jb} \cDb \p \Phib^{\jb}  &=
 - \Omega^-_{i\jb\kb} \p\Phib^{\jb} \cDb\Phib^{\kb}
 - \Omega^-_{ij\kb}  \p \Phi^j \cDb \Phib^{\kb} 
 - \cF^\alpha_{\beta\kb i} \cDb\Phib^{\kb} \Gammab_\alpha \Gamma^\beta, \nonumber\\
 \cDb \Gammab_\alpha & =0, \qquad \Gammab_\alpha \equiv \cH_{\alpha\alphab} \Gammab^{\alphab} .
\end{align}

\subsection{Symmetries of the classical action}
By construction the action is (0,2)-supersymmetric.  The action of the 
supercharges $\bQ$ and $\bQ$ on any superfield $X$ is defined by
\begin{align*}
 \sqrt{2} (\xi \bQ + \xib \bQb) \cdot X \equiv -\xi \cQ X -\xib \cQb X,
\end{align*}
where $\xi$ and $\xib$ denote constant Grassmann parameters. After
eliminating the auxiliary fields, the non-trivial transformations are as
follows:
\begin{align*}
\bQb \cdot \phib^{\ib}  &=-\psib^{\ib},  \quad 
\bQb \cdot \psi^i =\pb \phi^i;  \nonumber\\
\bQ \cdot \phi^i & = \psi^i, \quad
\bQ \cdot \psib^{\ib}  = -\pb\phib^{\ib}, \quad
\bQ \cdot \gamma^\beta  = -\cA^\beta_{\nu j} \psi^j\gamma^\nu, \quad
\bQ \cdot \gammab_{\alpha}  =  \cA^\nu_{\alpha j} \psi^j \gammab_\nu. 
\end{align*}
It is not hard to see that $\bQ^2 =\bQb^2 = 0$ and $\AC{\bQ}{\bQb} = \pb$;
the latter relation requires the use of the $\gamma$ equations of motion, while the former
hold off-shell.

It is also easy to see that corresponding to the $\GUL\times\GUR$ symmetries
we have the conserved currents $J_L = \gamma^\alpha\gammab_\alpha$, 
$J_R = g_{i\jb} \psi^i\psib^{\jb}$, satisfying $\pb J_L =0$ and $\p J_R =0$
up to equations of motion.  Similarly, we have the classical left-moving energy
momentum tensor
\begin{align}
\label{eq:classT}
T = -\frac{1}{\alpha'} \left\{g_{i\jb} \p\phi^i\p\phib^{\jb} 
+\frac{1}{2} (\gammab_\beta \p\gamma^\beta+\gamma^\beta\p\gammab_\beta)
+\cA^\mu_{\beta j} \p\phi^j \gammab_\mu\gamma^\beta\right\}.
\end{align}
$T$, like $J_L$, is annihilated by both $\bQ$ and $\bQb$ and hence conserved:
$\pb T = 0$.

\section{Massless $G$-neutral states via the (0,2) NLSM}\label{s:def}
If we assume that the NLSM describes a (0,2) SCFT, then we have all of the
tools necessary for constructing the massless spectrum of the corresponding
heterotic vacuum.  A typical approach is to determine the massless
fermions and infer the rest of the spectrum via  supersymmetry.  That is,
we work in the (NS,R) and (R,R) sectors of the theory and identify right-moving
ground states with $L_0$ eigenvalue of $+1$ for (NS,R) states and $L_0 = 0$ for
(R,R) states.  When working at tree-level
in the NLSM, it is possible to construct the states in the Born-Oppenheimer
approximation, where the mode expansion of the fields is truncated to right-moving
zero modes and first excited modes on the left~\cite{Rohm:1985jv}.  Working in this truncated
Fock space, we can then classify the states annihilated by $\bQ$ and $\bQb$
and having $L_0 =+1$.  Imposing the GSO projection, we will obtain the
tree-level spectrum of massless fermions.

The procedure sounds straightforward, and it would be surprising if it had
not already been applied to the (0,2) NLSM some time ago.  Indeed, the
computation is presented in~\cite{Distler:1987ee}, where
the massless spectrum is determined with one caveat:  \emph{``To be consistent,
we should include the first excited modes of [$\phi$], but as we are 
primarily interested in the gauge degrees of freedom, we will omit them.''}
That the excited modes of $\phi$ should contribute to the analysis is
reasonably clear, for instance from the last term in the classical energy-momentum
tensor in~(\ref{eq:classT}).  While this mixing
is indeed unimportant in the charged matter sector,\footnote{The (NS,R) charged matter states
involve a free left-moving fermion tensored with $\gamma$ or $\gammab$ and a wavefunction
of the bosonic zero modes;  there are no additional $\phi$ excitations.} it does 
affect the spectrum of neutral massless scalars arising from the (NS,R)
sector.

Our goal is to determine the neutral massless spectrum, keeping
track of all the necessary left-moving excitations.  However, instead of pursuing
the Born-Oppenheimer approach, we will attack the problem in a slightly different
fashion by studying equivalence classes of chiral operators in the NLSM.

\subsection{First order deformations of a (0,2) SCFT}
The $G$-neutral massless scalars of the four-dimensional effective theory have
a simple interpretation in the internal (0,2) SCFT as marginal $\GUL$-preserving 
first order deformations:  in the language of conformal perturbation theory, the
action is deformed by the integrated zero-momentum vertex operator for the
emission of the scalar.  The form of marginal supersymmetric deformations of
a unitary SCFT is tightly constrained.  For instance, in~\cite{Green:2010da} 
it is shown that in an $N=1$, $d=4$ superconformal theory the deformation must
be an F-term 
$$\Delta S = \int d^4 x ~d^2 \theta~ \cO + \text{h.c.},$$
where $O$ is a chiral primary operator with R-charge $2$;  there are
no non-trivial marginal D-term deformations.  A similar result holds in unitary (0,2)
SCFTs in two 
dimensions:\footnote{This is a consequence of  the (0,2) SCFT unitarity bounds~\cite{unpub:2dcft}.} 
a marginal supersymmetric deformation must take the form
\begin{align*}
\Delta S = \int d^2 z ~\cD X + \text{h.c.},
\end{align*}
where $X$ is a (0,2) chiral primary operator with $h=1$ and 
right-moving R-charge $\qb =+1$; as in the four-dimensional case,
a marginal deformation that is expressed as an integral over all of superspace
is necessarily trivial.

\subsection{Marginal superpotential deformations of the NLSM}
We will now assume that the (0,2) SCFT in question is well approximated by
a weakly coupled (0,2) NLSM.    Let $X$ be an operator in the SCFT of the type we 
just described.  Then in a classical (i.e. large radius) limit, $X$ must reduce
to $X_c$ --- a chiral superfield constructed from the NLSM fields with their classical
dimensions and charges listed in table~\ref{table:charges}.   
\begin{table}
\begin{center}
\begin{tabular}{|c|c|c|c|c|c|c|c|}

~		&$X$	&$\Gamma$	&$\Gammab$	&$\cD\Phi$	&$\cDb\Phib$	&$\p\Phi$	&$\p\Phib$ \\ \hline
$\qb$	&$1$		&$0$			&$0$			&$-1$		&$+1$		&$0$			&$0$ \\ \hline
$q$		&$0$		&$1$			&$-1$		&$0$			&$0$			&$0$			&$0$ \\ \hline
$\hb$	&$1/2$	&$0$			&$0$			&$1/2$		&$1/2$		&$0$			&$0$ \\ \hline
$h$		&$1$		&$1/2$		&$1/2$		&$0$			&$0$			&$1$			&$1$ \\ \hline
\end{tabular}
\caption{The classical charges and weights of NLSM fields}
\label{table:charges}
\end{center}
\end{table}
In other words, $X_c$ must be of the form
\begin{align*}
%\label{eq:genX}
X_c = \left[ \Gammab_{\alpha} \Gamma^\beta \Lambda^\alpha_{\beta \ib}(\Phi,\Phib) + \p \Phi^i Y_{i\ib}(\Phi,\Phib) 
+ \p\Phib^{\jb}  g_{i\jb} Z^i_{\ib}(\Phi,\Phib) \right]
\cDb \Phib^{\ib}.
\end{align*}
The NLSM fields are of course only defined in local coordinate patches, with
transition functions relating the fields in different patches. $X$ will be well-defined
across the patches if $\Lambda$, $Y$, and $Z$ take values in sections of certain bundles:
$$ \Lambda \in \Gamma( \End E \otimes \Omega^{0,1}_M), \qquad
     Y \in \Gamma(  \Omega^{1,1}_M), \qquad
     Z \in \Gamma( T_M \otimes \Omega^{0,1}_M),$$
where $\Omega_M^{p,q}$ denotes the (p,q) forms on the target-space $M$.

We have yet to impose that $X_c$ is chiral, i.e. $\cDb  X_c = 0$.  We need not require
$X_c$ to be chiral off-shell --- indeed, such a requirement would be too strong; instead,
as in~\cite{Beasley:2004ys,Beasley:2005iu}, we only require $\cDb X_c = 0$
up to the equations of motion of the unperturbed NLSM.  Computing $\cDb X_c$ and
using~(\ref{eq:eom}) to eliminate $\cDb \Gammab$ and $\p \cDb \Phib$ terms, we
obtain
\begin{align*}
\cDb X_c & = g_{i\jb } \p\Phib^{\jb} \cDb \Phib^{\kb} \cDb\Phib^{\ib} Z^i_{\ib,\kb}
+ \p\Phi^i\cDb\Phib^{\kb} \cDb\Phib^{\ib} ( Y_{i\ib,\kb} - H_{ji\kb} Z^j_{\ib} ) \nonumber\\
&\quad+ \Gammab_\alpha \Gamma^\beta \cDb\Phib^{\kb} \cDb\Phib^{\ib} 
(\Lambda^\alpha_{\beta \ib,\kb} - \cF^\alpha_{\beta \kb i} Z^i_{\ib} ).
\end{align*}
As there cannot be cancellations between the three terms, 
$\cDb X_c = 0$ requires
\begin{align}
\label{eq:cond}
Z^i_{\ib,\kb} - Z^i_{\kb,\ib} & = 0, \nonumber\\
Y_{i\ib,\kb} -Y_{i\kb,\ib} &= Z^j_{\kb} H_{ji\ib} - Z^j_{\ib} H_{ji\kb}, \\
\Lambda^\alpha_{\beta\ib,\kb}-\Lambda^\alpha_{\beta\kb,\ib} &= \cF^\alpha_{\beta\kb i} Z^i_{\ib} - \cF^\alpha_{\beta\ib i} Z^i_{\kb}. \nonumber
\end{align}
Of course not all solutions to~(\ref{eq:cond}) correspond to distinct first order deformation
of the SCFT --- a good thing, since the solution space is infinite dimensional; instead,
only certain equivalence classes of solutions correspond to deformations.  

To identify
the equivalence relations, we first consider another SCFT operator $X'$ with classical
limit $X'_c = X_c + \cDb W_c$ for some well-defined superfield $W_c$.  If
$X'$ and $X$ are distinct deformations of the theory, then their difference is a non-trivial deformation;  however, the latter would be a marginal deformation given as an integral 
over the full (0,2) superspace.  Since such deformations do not exist in the SCFT, we conclude
that $X$ and $X'$ define isomorphic deformations of the theory.  Conversely, if a
classical chiral superfield $X_c$ corresponds to a chiral primary operator $X$ in
the SCFT, then $X_c + \cDb W_c$ must correspond to the same first order deformation.

Thus, to count the first order deformations in the classical limit, we must consider
chiral superfields $X_c$ modulo the equivalence relation $X_c \sim X_c + \cDb W_c$.
In fact, there is another manner in which we can shift $X_c$ without affecting the
deformation: $X_c \to X_c + \p W'_c$ for some chiral superfield $\p W'_c$ leaves $\Delta S$ invariant.  As we will see, this additional equivalence will be trivial in most cases of interest.  So, to summarize, in the classical limit we expect the first order
deformations to correspond to $X_c$ that solve~(\ref{eq:cond}), modulo the equivalence relation
\begin{align*}
X_c \sim X_c + \cDb W_c + \p W'_c ,\qquad \cDb \p W' = 0.
\end{align*}
It is not difficult to make the equivalence more explicit --- we simply need to expand
$W_c$ and $W'_c$ in terms of the component fields.  Since we will now just work with
the classical NLSM Lagrangian, we will drop the $c$ subscripts on the fields.  Dimensional 
analysis and the $\GUL\times \GUR$ symmetry constrain $W$ and $W'$ to be
\begin{align*}
W =  \Gammab_{\alpha}\Gamma^\beta \lambda^\alpha_{\beta} + \p\Phi^i \mu_i +
 \p\Phib^{\ib}g_{i\ib} {\zeta}^i, \qquad
W'  = \cDb \Phib^{\ib}\xi_{\ib}, 
\end{align*}
where
$$
\lambda \in \Gamma(\End E ), \qquad
\mu \in \Omega^{1,0}_M, \qquad
\zeta \in \Gamma( T_M), \qquad \xi \in \Omega^{0,1}_M.
$$
$\p W'$ will be chiral up to the NLSM equations of motion provided that $\xi_{\ib}$ satisfies
\begin{align*}
\nabla^-_k \xi_{[\ib,\jb]} = 0, \qquad
\nabla^-_{\kb}\xi_{[\ib,\jb]} = 0, \qquad
g^{\jb i} (\cF^\alpha_{\beta\mb i} \xi_{[\ib,\jb]} - \cF^\alpha_{\beta\ib i} \xi_{[\mb,\jb]} ) = 0,
\end{align*}
where the $\nabla^-$ connection is defined with the twisted connection $\Omega^-$
given in~(\ref{eq:Omega}).  These conditions are solved by any $\pb$-closed $\xi$;
this is the most general solution for an $\SU(3)$ structure target-space~(see, e.g.~\cite{Grana:2005jc}).  O
therwise $\pb \xi$ would be a non-trivial $\nabla^-$-constant
form, in addition to the Hermitian form and the (3,0) form that define the $\SU(3)$ structure;  this would lead to 
a further reduction of structure.
Thus, we must have $\xi \in H^{0,1}_{\pb} (M)$. It is easy to see that when $\xi$ is cohomologically
trivial, it can be eliminated by redefining $\mu$ and $\zeta$.

Expanding out $\cDb W + \p W'$, we find the equivalence relation on $(\Lambda,Y,Z)$:
\begin{align}
\label{eq:equiv}
Z^i_{\ib} & \sim Z^i_{\ib} + (\zeta^i+g^{i\jb} \xi_{\jb})_{,\ib} +g^{i\kb}(\xi_{\ib,\kb}-\xi_{\kb,\ib}), \nonumber\\
Y_{i\ib} &\sim Y_{i\ib} +\mu_{i,\ib} + \xi_{\ib,i} + H_{ i\ib j}(\zeta^j+g^{j\jb} \xi_{\jb}) , \\
\Lambda^\alpha_{\beta\ib} & \sim
\Lambda^\alpha_{\beta\ib} +
\lambda^\alpha_{\beta,\ib} - \cF^\alpha_{\beta\ib i} (\zeta^i + g^{i\jb} \xi_{\jb}). \nonumber
\end{align}
Equations~(\ref{eq:cond}) and~(\ref{eq:equiv}) constitute our main result:  in a large radius limit the $G$-neutral first order deformations of a supersymmetric heterotic vacuum correspond to solutions of~(\ref{eq:cond}) modulo the equivalence relations in~(\ref{eq:equiv}).
The $Z$, $Y$ and $\Lambda$ are familiar from the textbook treatment of (2,2) compactifications and their deformations.  For instance, setting the right-hand sides of~(\ref{eq:cond}) to zero, we see that
$(Z,Y,\Lambda)$ define cohomology classes
$$ Z \in H^1(M, T_M), \qquad Y \in H^1( M,T^\ast_M), \qquad Z \in H^1(M,\End T_M).$$
However, the non-trivial right-hand sides indicate that for generic (0,2) theories the notion of splitting the deformations into ``complex structure, K\"ahler, and bundle'' is misleading.

\section{Examples} \label{s:examples}
Having obtained the general conditions, we can now check that they lead to the expected structure in
familiar limits of (2,2) theories and more general Calabi-Yau compactifications.  Having verified this, we
will be in a better position to discuss the implications for the general heterotic flux compactification.

\subsection{The (2,2) locus}

On the (2,2) locus, $E = T_M$, $H=0$ and $g$ is K\"ahler.   Moreover, since $H^{0,1}_{\pb} (M)$ is
trivial (we assume $M$ has the full $\SU(3)$ holonomy), we can set $\xi = 0$ without loss of generality.   
Thus, the equations reduce to
\begin{align}
\label{eq:22}
\pb Z & = 0, &   Z^i_{\ib} & \sim Z^i_{\ib} + \zeta^i_{,\ib}; \nonumber\\
\pb Y & = 0, &  Y_{i\ib} & \sim Y_{i\ib} + \mu_{i,\ib} ; \nonumber\\
\Lambda^m_{n\ib,\kb}  -
\Lambda^m_{n\kb,\ib} & = 
R^m_{~n\kb i} Z^i_{\ib} -R^m_{~n\ib i} Z^i_{\kb}, &
\Lambda^m_{n\ib} &\sim
\Lambda^m_{n\ib} + \lambda^m_{n,\ib} - R^m_{~n\ib i} \zeta^i,
\end{align}
where $R_{\mb n \kb i}$ is the Riemann tensor for the K\"ahler metric $g$.

As expected, deformations correspond to 
$Z \in H^1(M, T_M)$, $Y\in H^1(M,T^\ast_M)$; however, the conditions
on $\Lambda$ still appear a little bit puzzling.  The puzzle is easily resolved.
Let
$$ \Lambdat^m_{n\ib} \equiv \Lambda^m_{n\ib} - \nabla_n Z^m_{\ib}, \qquad
\lambdat^m_n \equiv \lambda^m_n - \nabla_n \zeta^m.$$
Recasting the last line of~(\ref{eq:22}) in terms of $\Lambdat$ and $\lambdat$,
we obtain
\begin{align}
\label{eq:22c}
\Lambdat^m_{n\ib,\kb}-\Lambdat^m_{n\kb,\ib} & = 
R^m_{~n\kb i} Z^i_{\ib} - R^m_{~n\ib i} Z^i_{\kb}
-(\nabla_n Z^m_{\ib})_{,\kb}
+(\nabla_n Z^m_{\kb})_{,\ib}, \\
\label{eq:22e}
\Lambdat^m_{n\ib} &\sim
\Lambdat^m_{n\ib}  + \lambdat^m_{n,\ib}
+ g^{m\mb} \left[\nabla_{\ib} \nabla_n \zeta_{\mb} -\nabla_n \nabla_{\ib} \zeta_{\mb} -  R_{\mb n\ib i} \zeta^i\right].
\end{align}
The square bracket in~(\ref{eq:22e}) is
$$\CO{\nabla_{\ib}}{\nabla_n} \zeta_{\mb} -R_{\mb n \ib i} \zeta^i
= (R_{\ib n\mb i}-R_{\mb n \ib i} )\zeta^i = 0,$$
where the last equality follows from the symmetry $R_{\ib n\mb i}=R_{\mb n \ib i}$
enjoyed by the Riemann tensor for a K\"ahler metric.  The vanishing of the
right-hand side of~(\ref{eq:22c}) follows from similar manipulations and $\pb Z = 0$.
Thus, in terms of the $\Lambdat$ and $\lambdat$ variables, we recover the
expected result: 
$$ \pb \Lambdat = 0, \qquad \Lambdat \sim \Lambdat + \pb \lambdat.$$
The first order deformations for a (2,2) compactification do have the canonical
split
$$ (Z,Y,\Lambdat)  \in H^1(M,T_M)\oplus H^1(M,T^\ast_M)\oplus H^1(M,\End T_M).$$

\subsection{Calabi-Yau compactifications}
A more generic (0,2) vacuum is obtained by taking $E\to M$ to be  a stable holomorphic bundle
over a (conformally) Calabi-Yau manifold.  In this case, the deformation space still has a 
familiar description.  Working at tree-level we still have $H=0$, and as in the (2,2) case
$\xi$ must be $\pb$-exact and hence can be absorbed into $\zeta$ and $\mu$.  Thus, $Z \in H^1(M,T_M)$, $Y\in H^1(M,T^\ast_M)$, and the remaining non-trivial
condition is
$$ \Lambda^\alpha_{\beta\ib,\kb} - \Lambda^\alpha_{\beta\kb,\ib} = 
 \cF^\alpha_{\beta\kb i} Z^i_{\ib} - \cF^\alpha_{\beta\ib i} Z^i_{\kb}.$$
Since $Z \in H^1(M,T_M)$ and $\cF$ is the (1,1) curvature for the holomorphic connection, the right-hand side defines a class in $H^2(M,\End E)$.  If this class
is trivial, then the equation can be solved for $\Lambda$; otherwise, the deformation is
obstructed.  As discussed at length in~\cite{Anderson:2011ty}, this is encoded in a long
exact sequence in cohomology~\cite{Atiyah:1957xx}, associated to the short
exact sequence
\begin{align*}
\xymatrix{ 0 \ar[r] & E\otimes E^\ast \ar[r] & Q \ar[r]^{\pi}  & T_M \ar[r] & 0 } ,
\end{align*}
\begin{align*}
\xymatrix{\cdots \ar[r] & H^1(M,Q) \ar[r]^{d\pi} &
 H^1(M,T_M) \ar[r]^-{\alpha} &H^2(M,E \otimes E^\ast) \ar[r] &\cdots},
\end{align*}
where the map $\alpha$ is given by contracting $Z \in H^1(M,T_M)$ with $\cF$.

\subsection{Application to heterotic flux vacua}
More generally, we hope to apply our results to heterotic compactifications on non-K\"ahler
manifolds.  These backgrounds are characterized by a tree-level $H$ background, the most
studied examples being $T^2$ bundles over K3~\cite{Dasgupta:1999ss,Becker:2006et,Becker:2009df}.  The NLSM $\alpha'$ expansion is rather formal for these backgrounds, as they generically contain string-scale cycles.  However, to the extent to which an $\alpha'$ expansion can be
used, our tree-level analysis describes the infinitesimal moduli of heterotic flux vacua.  The 
qualitative structure is quite sensible:  for instance,  the deformations of the complexified Hermitian form (the $Y_{i\ib}$) now have a non-trivial mixing with the complex structure deformations, and the ``breathing mode,'' corresponding to taking $Y$ proportional to the Hermitian form appears to be obstructed.  

It would be useful to clarify the geometry behind~(\ref{eq:cond}) and~(\ref{eq:equiv}).  For instance,
is it possible to prove that the space of these first-order deformations is finite dimensional for a smooth and compact
flux background?  Do $\SU(3)$ structure examples admit non-trivial $\xi$ equivalences?\footnote{Examples
with extended spacetime supersymmetry and hence reduced structure certainly possess
non-trivial $\xi$, see e.g.~\cite{Goldstein:2002pg}.}  How is this presentation of deformations related to 
the infinitesimal perturbations of solutions to the one-loop supergravity equations 
examined in~\cite{Becker:2006xp}?

\section{Concluding remarks} \label{s:discuss}
We have carried out the tree-level analysis of gauge-neutral massless scalars in a perturbative heterotic
vacuum based on a (0,2) NLSM.  Of course this is a far cry from providing a complete analysis of even
first-order deformations, let alone a picture of the (0,2) moduli space, and it is worthwhile to review the
limitations of our results.

First, our analysis has been carried out for compactifications based on $\SU(n)$ bundles over $M$ --- this
is the source of the $\GUL$ symmetry of the internal theory.  While this covers many vacua, it is certainly
not the most general situation, and there are certainly interesting compactifications based on $\GU(n)$
bundles, as well as more general constructions, e.g.~\cite{Blumenhagen:2005ga,Distler:2007av}.
Second, while it is natural (even technically so) to restrict to gauge-neutral scalars, at least as far as the
string perturbative limit is concerned, the Higgs deformations where $G$ is broken to some sub-group
should be considered on par with the neutral scalars we described.  Fortunately, at least the massless
charged spectrum has already been described in~\cite{Distler:1987ee}.

Modifications are also expected in going beyond tree-level in the  $\alpha'$ expansion.  In heterotic Calabi-Yau
compactifications the possible lifting of states is constrained by the axionic symmetries associated to shifts of the NLSM $B$-field~\cite{Green:1987sp,Polchinski:1998rr}; in more general heterotic flux compactifications
analogous constraints are not well understood.  At any rate, we certainly expect additional $G$-neutral massless
scalars associated to stringy enhanced symmetries, as well as lifting of states by world-sheet non-perturbative
effects.\footnote{Examples of these phenomena have recently been investigated in Landau-Ginzburg 
vacua~\cite{Aspinwall:2010ve,Aspinwall:2011us}.}

Although the general structure of deformations is complicated, since our analysis is just a simple application 
of (0,2) supersymmetry, it should be a good starting point for a systematic expansion in $\alpha'$ away from
the large radius limit.  For instance, it is reasonable
to expect that at one loop in $\alpha'$ the conditions will be modified by replacing $H$ with its gauge-invariant
form.  It would be interesting to see whether this expectation is borne out and to attempt to extend it to
an all orders result.

Other fruitful directions include applying these results to heterotic vacua with extended
spacetime supersymmetry (their NLSM description has been recently explored in~\cite{Melnikov:2010pq}),
as well relating them to gauged linear sigma model constructions.  The latter would be especially
interesting for the linear sigma models appropriate for flux backgrounds~\cite{Adams:2006kb,Blaszczyk:2011pg,Quigley:2011pg}.

%\bibliographystyle{utphys}
%\bibliography{bigref}
%\bibliographystyle{/Users/lmel/BIB/utphys}
%\bibliography{/Users/lmel/BIB/bigref}
%
\providecommand{\href}[2]{#2}\begingroup\raggedright\endgroup

\end{document}